# TITLE: CONSISTENT POLYNOMIAL EXPANSIONS OF STORED ENERGY FUNCTION FOR INCOMPRESSIBLE HYPERELASTIC MATERIALS

Aleksander FRANUS, aleksander.franus.dokt@pw.edu.pl, Faculty of Civil Engineering, Warsaw University of Technology, Aleja Armii Ludowej 16, 00-637 Warsaw, Poland

Stanisław JEMIOŁO, Faculty of Civil Engineering, Warsaw University of Technology

**Abstract** In the article, hyperelastic material models which state consistent polynomial expansions of the stored energy function are discussed. The approach follows from the multiplicative decomposition of the deformation gradient. Some advantages of the third order expansion model over the five-parameter Rivlin model using Treloar's experimental data are shown. The models are qualitatively and quantitatively compared to highlight these advantages of the discussed MV model.

## 1. Introduction

Elastomers are a group of polymers capable of very large elastic deformations. In a uniaxial tensile stress test, a deformation of a sample can reach 1000%. These include natural and synthetic rubbers. The capacity of elastomers to obtain large elastic deformations results from their internal structure [1]. Elastomers macromolecule chains form a complex spatial network, which may be described by statistical mechanics models. In this work, in order to describe some aspects of macroscopic properties of elastomers, the phenomenological approach resulting from mechanics of continuous media and theory of hyperelasticity of isotropic materials is considered [2].

Rubber-like materials are often described assuming their incompressibility [3]. There are many constitutive models proposed in the regime, which are derived from the phenomenological or the chain network approaches. In the literature since the 1940s, so called polynomial models have been proposed [4] [5], which are also widely used nowadays in engineering applications [6]. Formally, the most general Rivlin model for incompressible materials is a polynomial function of two basic invariants of isochoric deformation tensors. In order to reduce a number of material parameters, many simplifications of the Rivlin's proposal are postulated in the literature [7] [8]. Nevertheless, we do not discuss any type of non-polynomial models here.

This article is intended as an attempt to show that in the case of polynomial models for incompressible materials a consistent expansion of the stored energy function should be preferable to the Rivlin model. The idea follows from the multiplicative decomposition of the deformation gradient [9] rather than the expansion of the function such that only certain orders of the principal stretches are included [10] [11]. We demonstrate that the postulated so called MV model leads to qualitatively more accurate description of the classical Treloar's data than the discussed Rivlin model. It is especially noticed for large deformations, where the MV stored energy function exhibits desired regularity on the plane of principal stretches.

## 2. Basic equations of constitutive modelling

In this section basic equations concerning constitutive modelling of an isotropic, hyperelastic solid with finite deformations are presented [2] [12]. These show some aspects of the hyperelasticity which needs to be considered to correctly postulate a form of the stored energy function.

### 2.1. The stored energy function of isotropic hyperelastic materials

A deformation of an hyperelastic body is described by so called deformation gradient

$$\mathbf{F} \equiv \frac{\partial \chi(\mathbf{X},t)}{\partial \mathbf{X}}, \tag{1}$$

where $\mathbf{x} = \chi(\mathbf{X},t) = \mathbf{X} + \mathbf{u}(\mathbf{X},t)$ is a reversible vector function that defines the position of body particles in the current configuration ($\mathbf{u}$ stands for a displacement vector). In order to avoid interpenetration of matter, it is necessary to assume that $\mathbf{F} \in Lin^+$, where $Lin^+$ is a subset of $\mathbf{F} \in Lin$ (a nine dimensional space of second order tensors) such that $\det \mathbf{F} > 0$ [2]. Then, there exists a unique polar decomposition

$$\mathbf{F} = \mathbf{RU} = \mathbf{VR}, \tag{2}$$

where $\mathbf{R}$ is an orthogonal rotation tensor and $\mathbf{U}$, $\mathbf{V}$ are positive definite stretch tensors such that

$$\mathbf{U} = \sqrt{\mathbf{C}} = \sqrt{\mathbf{F}^T \mathbf{F}}, \qquad \mathbf{V} = \sqrt{\mathbf{B}} = \sqrt{\mathbf{F}\mathbf{F}^T}. \tag{3}$$

Tensors $\mathbf{C}$ and $\mathbf{B}$ are the left and the right Cauchy-Green deformation tensors, respectively.

In the theory of hyperelasticity it is assumed an existence of a sufficiently regular function $W$ called stored energy function so that

$$\mathbf{S}(\mathbf{F}) = \frac{\partial W(\mathbf{F})}{\partial \mathbf{F}}, \tag{4}$$

where $\mathbf{S}$ denotes the first Piola-Kirchhoff stress tensor. However, restrictions imposed by objectivity require that the stored energy should not be directly described only by the deformation gradient [12] [3]. Under the assumption of isotropy of the stored energy function, it yields

$$W(\mathbf{Q}\mathbf{C}\mathbf{Q}^T) = W(\mathbf{C}), \quad \forall \mathbf{Q} \in O(3). \tag{5}$$

The stored energy function may be also described in terms of the stretch tensor $\mathbf{U}$.

The isotropic scalar function $W(\mathbf{C})$ may be rewritten as a function of three invariants of tensor $\mathbf{C}$ and $\mathbf{B}$ such that

$$W(\mathbf{C}) = \hat{W}(\mathrm{tr}\mathbf{C}, \mathrm{tr}\mathbf{C}^2, \mathrm{tr}\mathbf{C}^3) = \hat{W}(\mathrm{tr}\mathbf{B}, \mathrm{tr}\mathbf{B}^2, \mathrm{tr}\mathbf{B}^3) = \breve{W}(I_1, I_2, I_3), \tag{6}$$

where the basic invariants are functions of $\mathbf{C}$ and $\mathbf{B}$ as follows

$$I_1 = \breve{I}_1(\mathbf{B}) = \breve{I}_1(\mathbf{C}) = \mathrm{tr}\mathbf{C}, \quad I_2 = \breve{I}_2(\mathbf{B}) = \breve{I}_2(\mathbf{C}) = \frac{1}{2}\left[(\mathrm{tr}\mathbf{C})^2 - \mathrm{tr}\mathbf{C}^2\right],$$
$$I_3 = \breve{I}_3(\mathbf{B}) = \breve{I}_3(\mathbf{C}) = \det\mathbf{C} = \frac{1}{3}\left(\frac{1}{2}(\mathrm{tr}\mathbf{C})^3 - \frac{3}{2}\mathrm{tr}\mathbf{C}\,\mathrm{tr}\mathbf{C}^2 + \mathrm{tr}\mathbf{C}^3\right). \tag{7}$$

The set of invariants $\{\mathrm{tr}\mathbf{C}, \mathrm{tr}\mathbf{C}^2, \mathrm{tr}\mathbf{C}^3\}$ states a polynomial, irreducible basic of isotropic function [3]. The invariants are symmetric functions of the eigenvalues $\lambda_k > 0$ of the stretch tensors as follows

$$I_1 = \lambda_1^2 + \lambda_2^2 + \lambda_3^2, \quad I_2 = \lambda_2^2\lambda_3^2 + \lambda_3^2\lambda_1^2 + \lambda_1^2\lambda_2^2, \quad I_3 = \lambda_1^2\lambda_2^2\lambda_3^2 = J^2. \tag{8}$$

Thus, the stored energy function may be also expressed as a symmetric function of the eigenvalues $\lambda_k$ such that [2]

$$W(\mathbf{U}) = W(\mathbf{V}) = \tilde{W}(\lambda_1, \lambda_2, \lambda_3) = \tilde{W}(\lambda_2, \lambda_3, \lambda_1) = \tilde{W}(\lambda_3, \lambda_1, \lambda_2) =$$
$$= \tilde{W}(\lambda_1, \lambda_3, \lambda_2) = \tilde{W}(\lambda_3, \lambda_2, \lambda_1) = \tilde{W}(\lambda_2, \lambda_1, \lambda_3). \tag{9}$$

A crucial role in the theory of hyperelasticity plays polyconvexity as it ensures existence of an associated boundary value problem [13]. A function $W$ is polyconvex if there exists a convex function $\mathcal{W}$ such that

$$W(\mathbf{F}) = \mathcal{W}(\mathbf{F}, \mathrm{Cof}\mathbf{F}, \det\mathbf{F}), \tag{10}$$

with $(\mathbf{F}, \mathrm{Cof}\mathbf{F}, \det\mathbf{F}) \in Lin \times Lin \times (0, \infty)$. We emphasize that $\mathbf{F}, \mathrm{cof}\,\mathbf{F}, \det\mathbf{F}$ concern a deformation of lines, a surface and a volume of the body, respectively. The extended function is constructed that way mostly because the set $Lin^+$ is not convex. Moreover, it is requaired to put some other restrictions on $\mathcal{W}$ to ensure physically motivated proper growth conditions.

## 2.2. Constitutive relations for incompressible solids

In the case of incompressibility, we assume that a material is subjected to internal constraints of the form

$$J - 1 = \det \mathbf{F} - 1 = 0, \tag{11}$$

which ensure lack of volumetric changes of a body [14]. It might be rewritten as

$$\det \mathbf{C} = \frac{1}{3}\left(\frac{1}{2}(\mathrm{tr}\mathbf{C})^3 - \frac{3}{2}\mathrm{tr}\,\mathbf{C}\,\mathrm{tr}\,\mathbf{C}^2 + \mathrm{tr}\,\mathbf{C}^3\right) = 1, \tag{12}$$

which implies directly that the trace of the third power of tensor $\mathbf{C}$ is reducible. However, it is convenient to express the stored energy function in terms of two modified invariants of $\bar{\mathbf{C}} = \bar{\mathbf{F}}^T\bar{\mathbf{F}}$. It follows from the decomposition of the deformation gradient $\mathbf{F} = J^{1/3}\bar{\mathbf{F}}$ so that the volumetric and isochoric deformations are decoupled [9]. The formulation is particularly convenient when describing slightly compressible material [15] [16]. Here, the stored energy function is postulated by

$$W(\bar{\mathbf{C}}) = \hat{W}\left(\mathrm{tr}\bar{\mathbf{C}}, \mathrm{tr}\bar{\mathbf{C}}^2\right) = \breve{W}\left(\bar{I}_1, \bar{I}_2\right), \tag{13}$$

where

$$\bar{I}_1 = \mathrm{tr}\bar{\mathbf{C}} = \mathrm{tr}\bar{\mathbf{B}}, \quad \bar{I}_2 = \frac{1}{2}\left[(\mathrm{tr}\bar{\mathbf{C}})^2 - \mathrm{tr}\bar{\mathbf{C}}^2\right] = \mathrm{tr}\bar{\mathbf{C}}^{-1} = \mathrm{tr}\bar{\mathbf{B}}^{-1}. \tag{14}$$

$W(\bar{\mathbf{F}})$ is polyconvex if there exists an extension of the function to convex one $\mathcal{W}$ such that $W(\bar{\mathbf{F}}) = \mathcal{W}(\mathbf{F}, \mathrm{cof}\mathbf{F})$.

The function $\breve{W}(\bar{I}_1, \bar{I}_2)$ is not an elastic potential, because the volumetric part of the Cauchy stress tensor is not specified. The constitutive relation, in the deformed body configuration, defines only the deviatoric part of the Cauchy stress

$$\boldsymbol{\sigma} = -p\mathbf{I} + 2\frac{\partial \breve{W}\left(\bar{I}_1, \bar{I}_2\right)}{\partial \bar{I}_1}\operatorname{dev}\left(\bar{\mathbf{B}}\right) - 2\frac{\partial \breve{W}\left(\bar{I}_1, \bar{I}_2\right)}{\partial \bar{I}_2}\operatorname{dev}\left(\bar{\mathbf{B}}^{-1}\right) , \qquad (15)$$

where $\operatorname{dev}(\bullet)$ denotes the deviatoric component of a tensor. The hydrostatic pressure component $p$ does not depend on the particular material definition.

## 3. Consistent polynomial expansions of the stored energy function

Constitutive models for isotropic incompressible hyperelastic materials are continuously proposed and discussed in the literature [17] [18] [19] [7] [20]. A representation of the stored energy function is typically considered as follows:

- the arguments of the stored energy function are invariants of the polynomial basis,
- the arguments of the stored energy function are the eigenvalues of the stretch tensors,
- the arguments of the stored energy function are other three independent invariants of any reasonable tensor measure of deformation or strain.

From the point of view of the theory of representation of tensor functions, the choice of a functional basis other than $\{\operatorname{tr}\mathbf{C}, \operatorname{tr}\mathbf{C}^2, \operatorname{tr}\mathbf{C}^3\}$ is not theoretically important [3] but might be more convenient for studying a certain computational aspects. An important distinction between models is their linearity or non-linearity in terms of model parameters. All of the above statements also hold in the case of the incompressibility.

In the article we focus on showing a construction of polynomial models for incompressible rubber-like materials which follows from a consistent expansion of the stored energy function in terms of the measure of deformation $\bar{\mathbf{C}}$. In the opinion of the authors, the discussed MV model is the simplest polynomial one which captures a typical 'S' shape of a stress-stretch curve of rubber and gives a sufficiently regular stored energy function for large deformations.

A similar idea of consistent expansion in the context of stress-stretch relationships resulting from basic homogeneous deformations and their importance for the description of elastomers are discussed e.g. in the articles [21] [22] [23] [10]. The basic motivation given in the papers is the description of the 'S' shape of a stress-stretch curve, which is not predicted by the Mooney model [4]. Therefore, it was proposed to include only certain orders of principal stretches, i.e. $\lambda^2, \lambda^4, \lambda^6$, in an expansion the stored energy function [11]. The reasoning is not correct, because the incompressibility assumption $J = \lambda_1 \lambda_2 \lambda_3 = 1$ leads to $\lambda_3 = 1/(\lambda_1 \lambda_2)$. Nevertheless, these relations were successfully applied in the works to approximate experimental data of elastomers.

Here, the consistent $n^{\text{th}}$ order polynomial expansion $\hat{W}(\operatorname{tr}\bar{\mathbf{C}}, \operatorname{tr}\bar{\mathbf{C}}^2)$ of the stored energy function $W(\bar{\mathbf{C}})$ fulfills the condition

$$\left| W(\bar{\mathbf{C}}) - \hat{W}(\operatorname{tr}\bar{\mathbf{C}}, \operatorname{tr}\bar{\mathbf{C}}^2) \right| = \mathcal{O}\left( \|\bar{\mathbf{C}}\|_2^{n+1} \right) \tag{16}$$

where $\|\bullet\|_2$ denotes the Euclidian norm. We restive ourselves to at most third order terms $(n=3)$. Hence, a general form of $\hat{W}$ yields

$$\hat{W}(\operatorname{tr}\bar{\mathbf{C}}, \operatorname{tr}\bar{\mathbf{C}}^2) = b + c\operatorname{tr}\bar{\mathbf{C}} + d_1 (\operatorname{tr}\bar{\mathbf{C}})^2 + d_2 \operatorname{tr}\bar{\mathbf{C}}^2 + e_1 (\operatorname{tr}\bar{\mathbf{C}})^3 + e_2 \operatorname{tr}\bar{\mathbf{C}} \operatorname{tr}\bar{\mathbf{C}}^2, \tag{17}$$

Assuming that the reference configuration is a natural state

$$\hat{W}(3,3) = b + 3c + 9d_1 + 9d_2 + 27e_1 + 27e_2 = 0, \tag{18}$$

we obtain the function of the form

$$\hat{W}(\operatorname{tr}\bar{\mathbf{C}}, \operatorname{tr}\bar{\mathbf{C}}^2) = c(\operatorname{tr}\bar{\mathbf{C}} - 3) + d_1\left[(\operatorname{tr}\bar{\mathbf{C}})^2 - 9\right] + d_2(\operatorname{tr}\bar{\mathbf{C}}^2 - 9) + \\ + e_1\left((\operatorname{tr}\bar{\mathbf{C}})^3 - 27\right) + e_2(\operatorname{tr}\bar{\mathbf{C}} \operatorname{tr}\bar{\mathbf{C}}^2 - 27). \tag{19}$$

Substituting definitions of the invariants into (16) yields five-parameter MV model [3]

$$\hat{W}\left(\operatorname{tr}\bar{\mathbf{C}},\operatorname{tr}\bar{\mathbf{C}}^2\right)=\breve{W}_{MV}\left(\bar{I}_1,\bar{I}_2\right)=$$
$$=\frac{1}{2}\left[a_1\left(\bar{I}_1-3\right)+\frac{1}{2}a_2\left(\bar{I}_1^2-9\right)+\frac{1}{3}a_3\left(\bar{I}_1^3-27\right)+a_4\left(\bar{I}_2-3\right)+a_5\left(\bar{I}_1\bar{I}_2-9\right)\right]. \tag{20}$$

More strictly speaking, the stored energy function (20) is consistent third order expansion in terms of the deformation measure $\bar{\mathbf{C}}$. Nevertheless, the polyconvexity of the function is not ensured because of the term $\bar{I}_1\bar{I}_2$.

Analogously, as a special case of (20), we obtain the stored energy function of the Ishihara-Zahorski (MIZ) model [24] [25]

$$\breve{W}_{MIZ}\left(\bar{I}_1,\bar{I}_2\right)=\frac{1}{2}\left[a_1\left(\bar{I}_1-3\right)+\frac{1}{2}a_2\left(\bar{I}_1^2-9\right)+a_4\left(\bar{I}_2-3\right)\right], \tag{21}$$

which states the second order model in the sense of (16). The function (20) may be rewritten in a slightly different form

$$\breve{W}_{MIZ}\left(\bar{I}_1,\bar{I}_2\right)=\frac{1}{2}\mu_0\left[\alpha\left(\bar{I}_1-3\right)+(1-\alpha)\left(\bar{I}_2-3\right)\right]+\frac{1}{4}c\left(\bar{I}_1-3\right)^2, \tag{22}$$

where $\mu_0>0$ denotes an initial shear modulus. The expression (21) is not as general as (22), but more convenient for analysing some aspects of non-uniqueness of a boundary value problem solution [26]. The function (21) is polyconvex if material parameters satisfy $\alpha\in(0,1)$ and $c>0$ [19]. For $\alpha=1, c=0$ we formally obtain the neo-Hookean stored energy function

$$\breve{W}_{NH}\left(\bar{I}_1,\bar{I}_2\right)=\frac{1}{2}\mu_0\left(\bar{I}_1-3\right) \tag{23}$$

which is the simplest constitutive model for incompressible materials and states the first order expansion in the sense of (16).

It is worth noticing that the other well-known in the literature polynomial models follow from (19). Assumption $a_4=a_5=0$ yields the Yeoh model [27] which is given by the function of the first invariant only. Adding to the Yeoh form a linear term of the second invariant gives the Biderman model [28]. None of these states a consistent expansion of the stored energy function in terms of $\bar{\mathbf{C}}$.

## 4. Comparison of the MV model with the Rivlin polynomial form

In the following part of the paper, we compare a performance of presented models. In particular, we demonstrate some advantages of the MV model in comparison with the five-parameter Rivlin model on the basis of classical Treloar's experimental data on a vulcanized rubber [29]. The method of least squares is employed to minimize the sum of the squares of the relative error

$$\arg\min_{\mathbf{a}\in\mathbb{R}^m} \sum_{i=1}^{n}\left[1 - \frac{S_j^{(i)}\left(\mathbf{a}, \bar{\lambda}_j^{(i)}\right)}{S_j^{(i)}}\right]^2, \quad (24)$$

where $\mathbf{a}$ stands for a set of a model parameters. As a measure of the quality of fit we choose an estimated variance of residuals $r^{(i)}$ such that

$$s^2 = \frac{1}{n-p}\sum_{i=1}^{n}\frac{1}{\left(\sigma_j^{(i)}\right)^2}\left(r^{(i)}(\mathbf{a})\right)^2, \quad (25)$$

where $p$ denotes the number of parameters in a model and $n$ is the number of data elements. For more details on the subject of parameters identification for hyperelastic models we refer the reader to [30] [31].

### 4.1. The MV model

As concluded in the previous section, the neo-Hookean model, the MIZ model and the MV model are the first, the second and the third consistent expansions of the stored energy function in the sense of (16), respectively. It is known that the neo-Hookean model poorly describes elastomers mostly due to lack of influence of the second invariant. However, it can be applied to model these type of materials in a very small range of deformation, see Fig.1. It is known that the MIZ model brings qualitatively and quantitively better results in comparison with the neo-Hookean model. In Fig. 2a it is shown that the model accurately describes the data but only for moderate deformations. The whole range of deformations is properly approximated by the MV model, see Fig 2b. However, evaluating a model fit based on an accuracy of fitting a curve to data is not sufficient, because the determined parameters should not only to reasonably reproduce the material behavior in the selected basic tests, but to allow a reasonable description of an arbitrary deformation. In this context even a cursory evaluation of contour plots of a stored energy function allows rejecting completely inadequate sets of parameters. Hence, it should be emphasized that contour plots of stored energy functions for all three models are very regular, see Fig. 3. In the following part of the article, we demonstrate that it is not the case for the five-parameter Rivlin model.

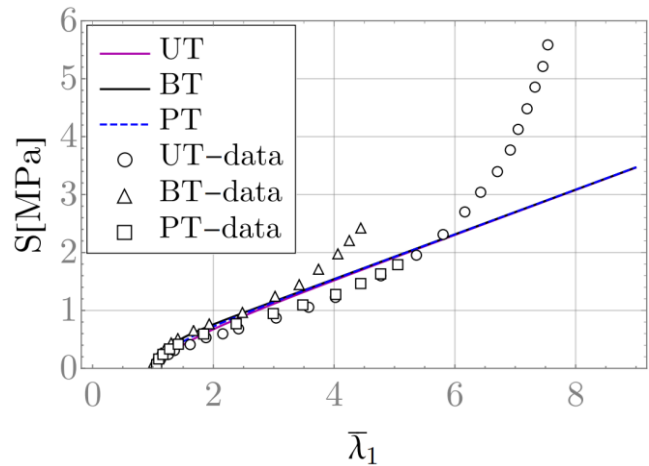

**Fig. 1.** Plots of the nominal stress vs principal stretch – the neo-Hookean model.

It is worth mentioning here that the second invariant $\bar{I}_2$ is not a convex function for sufficiently large stretches [19]. However, the stored energy function for basic experimental tests can be considered as a function of a single variable $W(\bar{\lambda}_1)$. Then, the functions $\bar{I}_1(\bar{\lambda}_1), \bar{I}_2(\bar{\lambda}_1)$ are convex.

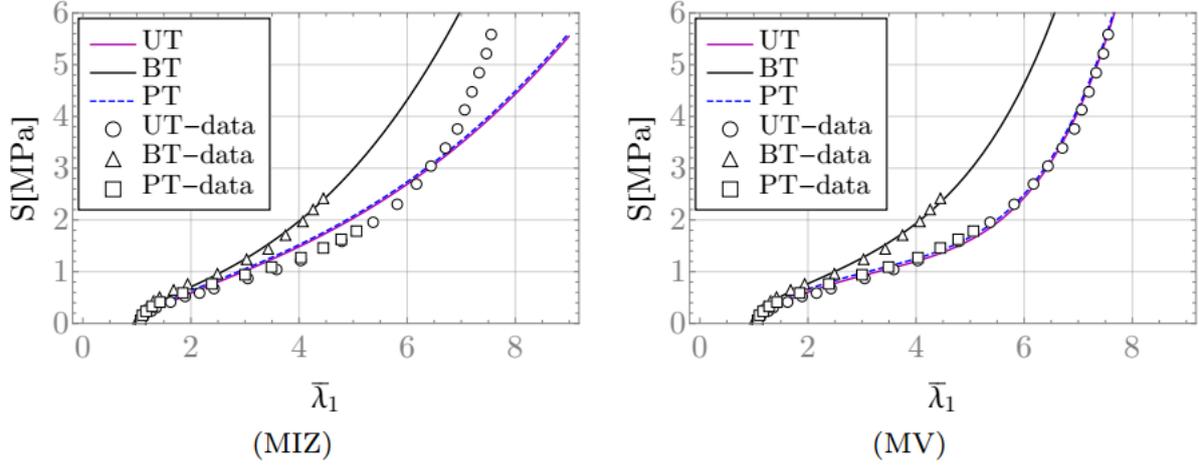

**Fig. 2.** Plots of the nominal stress vs principal stretch – the MIZ and MV models.

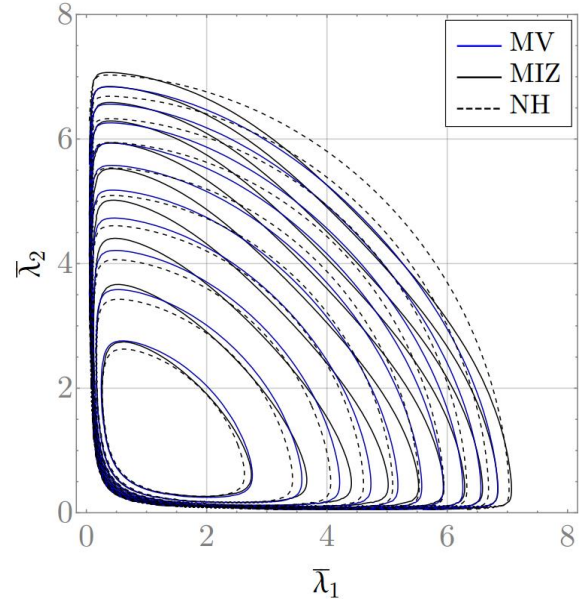

**Fig. 3.** Contour plots of stored energy function of the neo-Hookean (NH), the MIZ and the MV models.

Nonconvexity of the second invariant $\bar{I}_2(\bar{\lambda}_1, \bar{\lambda}_2)$ might lead to nonconvexity of a stored energy function with respect to the principal stretches [12]. As an example, we utilize the MIZ model whose parameters should satisfy $\alpha \in (0,1)$ and $c > 0$ to ensure polyconvexity of the stored energy function. For $f = 0.8, c = 0.1$ the function $\tilde{W}(\bar{\lambda}_1, \bar{\lambda}_2)$ is convex, which is not the case for $f = 0.2, c = 0.1$, see Fig. 4.

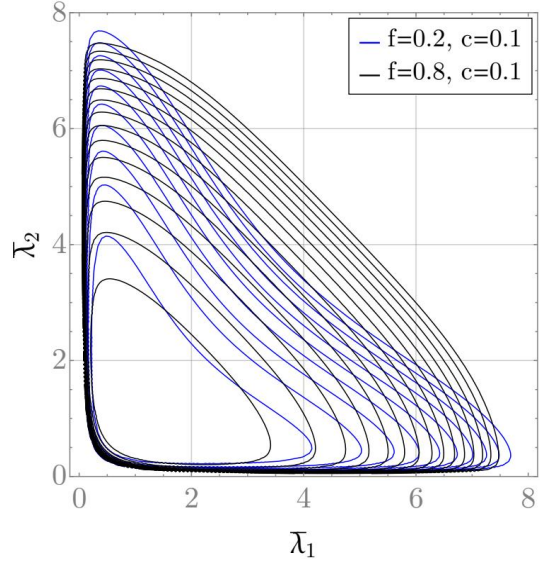

**Fig. 4.** Contour plots of the stored energy function of the MIZ model showing non-convexity of the function for certain set of parameters.

### 4.2. The Rivlin polynomial model

Rivlin [5] proposed a phenomenological model of the isotropic, hyperelastic incompressible material of the form

$$U\left(\bar{I}_1, \bar{I}_2\right) = \sum_{k+l=1}^{N} C_{kl} \left(\bar{I}_1 - 3\right)^k \left(\bar{I}_2 - 3\right)^l, \qquad (25)$$

where $N$ is an arbitrary natural number. Formally, the polynomial (12) can be seen as a series expansion of the stored energy function at the point $\bar{I}_1 = 3, \bar{I}_2 = 3$ such that

$$\begin{aligned}
U\left(\bar{I}_1, \bar{I}_2\right) &= U(3,3) + \frac{\partial U(3,3)}{\partial \bar{I}_1}\left(\bar{I}_1 - 3\right) + \frac{1}{2}\frac{\partial^2 U(3,3)}{\partial \bar{I}_1^2}\left(\bar{I}_1 - 3\right)^2 + \\
&+ \frac{\partial U(3,3)}{\partial \bar{I}_2}\left(\bar{I}_2 - 3\right) + \frac{1}{2}\frac{\partial^2 U(3,3)}{\partial \bar{I}_2^2}\left(\bar{I}_2 - 3\right)^2 + \frac{\partial^2 U(3,3)}{\partial \bar{I}_1 \partial \bar{I}_2}\left(\bar{I}_1 - 3\right)\left(\bar{I}_2 - 3\right) + \ldots,
\end{aligned} \qquad (26)$$

A motivation to compare the model with the MV is its wide popularity in applications [32] [33]. For example, in the documentation of the ABAQUS [6] it is stated that the model for $N=2$ more accurately approximates the Treloar's data than the Mooney model, i.e. $N=1$. However, the comparison is performed based on the quality of fit to the data which describe only certain deformation modes. We show that the five-parameter model

$$\breve{W}_{MR}\left(\bar{I}_1,\bar{I}_2\right) = \sum_{k+l=1}^{N=2} C_{kl}\left(\bar{I}_1-3\right)^k\left(\bar{I}_2-3\right)^l = C_{10}\left(\bar{I}_1-3\right) + C_{20}\left(\bar{I}_1-3\right)^2 + \\ + C_{01}\left(\bar{I}_2-3\right) + C_{02}\left(\bar{I}_2-3\right)^2 + C_{11}\left(\bar{I}_1-3\right)\left(\bar{I}_2-3\right)$$

(27)

performs qualitatively much poorer than the MIZ and MV models for Treloar's data.

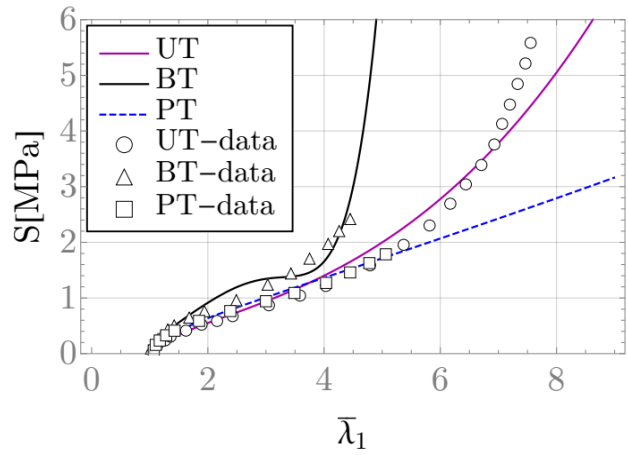

**Fig. 5.** Plots of the nominal stress vs principal stretch – the five-parameter Rivlin model.

Comparing the results based on the estimated variance in Tab.1, we do not notice a significant difference between these three models. In fact, the variance for the Rivil model is lower than for the MIZ one. Nevertheless, the approximation of the data is not qualitatively acceptable, especially for large deformations, see Fig. 5. More importantly, contour plots of the stored energy function with respect to the principal stretches show extensive irregularity of the function. There is no physical justification of such behavior of the elastomer. It states a serious disadvantage of the model in the case of simulating large non-homogenous deformations e.g. with a help of the finite element method [14].

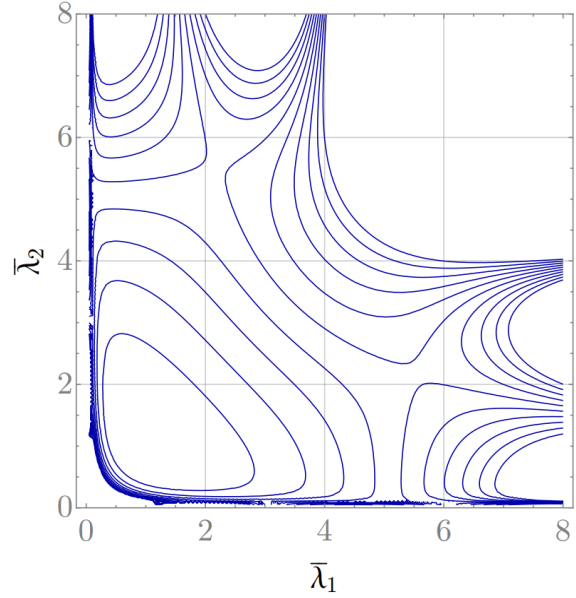

**Fig. 6.** Contour plots of stored energy function $\breve{W}_{MR}(\bar{I}_1, \bar{I}_2)$ of the five-parameter Rivlin model.

**Table 1.** Parameters for the Treloar's data

| Model | $a_1$ | $a_2$ | $a_3$ | $a_4$ | $a_5$ | Variance |
|---|---|---|---|---|---|---|
| MIZ | $3.139 \times 10^{-1}$ | $3.746 \times 10^{-3}$ | - | $3.789 \times 10^{-3}$ | - | $4.023 \times 10^{-2}$ |
| MV | $3.735 \times 10^{-1}$ | $-8.634 \times 10^{-3}$ | $2.644 \times 10^{-4}$ | $2.078 \times 10^{-2}$ | $-2.825 \times 10^{-4}$ | $1.932 \times 10^{-2}$ |
| Model | $C_{01}$ | $C_{10}$ | $C_{02}$ | $C_{20}$ | $C_{11}$ | Variance |
| Rivlin | $3.164 \times 10^{-2}$ | $1.383 \times 10^{-1}$ | $9.034 \times 10^{-5}$ | $1.716 \times 10^{-3}$ | $-1.769 \times 10^{-3}$ | $3.026 \times 10^{-2}$ |

## 5. Concluding remarks

The MV model, which appears to be the consistent third order polynomial expansion of the stored energy function, describes the Treloar's data on rubber more accurately the presented polynomial Rivlin model. Its advantage is especially noticed in a range of large deformations. Two-dimensional contours plots of the stored energy function on the principal stretches plane exhibits its desired regularity. It is not the case for discussed the five-parameter Rivlin model.

An evaluation of a stored energy function based on contours plots on the principal stretches plane allows to reject inadequate sets of parameters or even a model [3]. An accuracy of the stored energy function fit should not be only conducted based on specific deformation modes corresponding to the experimental data. The model should be capable to predict a reasonable response to any deformation. A similar approach for comparing stored energy functions in the plane of the isochoric invariant was developed in [34], [35]. The approach might be also useful for an evaluation of other type of the stored energy function, but it certainly would not reveal the irregularity shown in the case of the Rivlin model.